\documentclass{revtex4}
\usepackage{amssymb}
\usepackage{amsmath}
\usepackage{graphicx}
\usepackage[font={footnotesize,it}]{caption}

\pdfoutput=1 
\input{tcilatex}

\begin{document}

\title{Charge screening by thin-shells in a 2+1-dimensional regular black
hole }
\author{S. Habib Mazharimousavi}
\email{habib.mazhari@emu.edu.tr}
\author{M. Halilsoy}
\email{mustafa.halilsoy@emu.edu.tr}
\affiliation{Physics Department, Eastern Mediterranean University, G. Magusa north
Cyprus, Mersin 10 Turkey}

\begin{abstract}
We consider a particular Bardeen black hole in 2+1-dimensions. The black
hole is sourced by a radial electric field in non-linear electrodynamics
(NED). The solution is obtained anew by the alternative Hamiltonian
formalism. For $r\rightarrow \infty $ it asymptotes to the charged BTZ black
hole. It is shown that by inserting a charged, thin-shell (or ring) the
charge of the regular black hole can be screened from the external world.
\end{abstract}

\maketitle

\section{Introduction}

Charge is one of the principal hairs associated with black holes that can be
detected classically / quantum mechanically by external observers. The
question that naturally may arise is the following: By some artefact is it
possible to hide charge from distant observers? This is precisely what we
aim to answer in a toy model of a regular Bardeen black hole in $2+1-$%
dimensions. For this purpose we revisit a known black hole solution powered
by a source from nonlinear electrodynamics (NED) \cite{1}. With the advent
of NED coupled to gravity interesting solutions emerge as a result. The
reason for this richness originates from the arbitrary self-interaction of
electromagnetic field paving the way to a large set of possible Lagrangians.
From its inception NED has built a good reputation in removing singularities
due to point charges \cite{2}. This curative power of NED can equally be
adopted to general relativity where spacetime singularities play a prominent
role. As an example we cite the Reissner-Nordstrom (RN) solution which is
known to suffer from the central, less harmful time-like singularity. By
replacing the linear Maxwell theory with NED it was shown that the spacetime
singularity can be removed \cite{3,4}. For similar purposes NED can be
employed in different theories as well. Let us add that one should not
conclude that all gravity-coupled NED solutions are singularity free. For
instance, we gave newly a solution in $2+1-$dimensions where the Maxwell's
field tensor is $F_{\mu \nu }=E_{0}\delta _{\mu }^{t}\delta _{\nu }^{\theta }
$, ($E_{0}=$constant), which yields a singular solution \cite{6}.

We must also add that apart from introducing NED coupling to make a regular
RN an alternative approach was considered long ago by Israel \cite{5}. In 
\cite{5} it was considered a collapsing spherical shell as a source for the
Einstein-linear Maxwell theory which served equally well to remove the
central singularity.

In this paper we elaborate on a regular Bardeen black hole in $2+1-$%
dimensions \cite{1}. We rederive it by applying a Legendre transformation so
that from the Lagrangian we shift to Hamiltonian of the system. The
Lagrangian of the involved NED model turns out to be transcendental whereas
the Hamiltonian becomes tractable. We show that at least the weak energy
conditions (WECs) are satisfied. By applying the extrinsic curvature
formalism of Lanczos (i.e. the cutting and pasting method) \cite{7} we match
the regular interior to the chargeless BTZ spacetime \cite{8} outside. The
boundary in between is a stable thin-shell, (or intrinsically a ring) which
is the trivial version of an FRW universe. The choice of charge on the
thin-shell with appropriate boundary conditions renders outside to be free
of charge. This amounts, by construct, to shield inner charge of the
spacetime (herein a Bardeen black hole) from the external observer. The idea
can naturally be extended to higher dimensional spacetimes to eliminate
black hole's charge, or other hairs by artificial setups.

The paper is organized as follows. In Sec. II we derive the Bardeen black
hole from the Hamiltonian formalism; the energy conditions and simple
thermodynamics are presented. Charge screening effect and stability of
thin-shell are described in Sec. III. The paper is completed with Conclusion
in Sec. IV.

\section{Bardeen Black Hole in $2+1-$dimensions}

\subsection{Rederivation of the solution using Hamiltonian method}

Bardeen's black hole in $2+1-$dimensions was found by Cataldo et al. \cite{1}%
. They represented a regular black hole in $2+1-$dimensions whose source, in
analogy with $3+1-$dimensional counterpart \cite{3}, is NED. In this section
first we revisit the solution by introducing the Hamiltonian of the system.
The $2+1-$dimensional action reads 
\begin{equation}
I=\frac{1}{2}\int dx^{3}\sqrt{-g}\left( R-2\Lambda -\mathcal{L}\left( 
\mathcal{F}\right) \right) 
\end{equation}%
in which $\mathcal{F}=F_{\mu \nu }F^{\mu \nu }$ is the Maxwell invariant
with $R$ the Ricci scalar and $\Lambda $ the cosmological constant. The line
element is circular symmetric written as 
\begin{equation}
ds^{2}=-A\left( r\right) dt^{2}+\frac{dr^{2}}{A\left( r\right) }%
+r^{2}d\theta ^{2},
\end{equation}%
where $A(r)$ is the metric function to be determined. The field $2-$form is
chosen to be pure radial electric field (as in the charged BTZ) 
\begin{equation}
\mathbf{F}=E\left( r\right) dt\wedge dr
\end{equation}%
in which $E\left( r\right) $ stands for the electric field to be found. The
Maxwell's nonlinear equation is%
\begin{equation}
d\left( ^{\star }\mathbf{F}\frac{\partial \mathcal{L}}{\partial \mathcal{F}}%
\right) =0,
\end{equation}%
where $^{\star }\mathbf{F}$ is the dual of $\mathbf{F}$ while the
Einstein-NED equations read 
\begin{equation}
G_{\mu }^{\nu }+\Lambda g_{\mu }^{\nu }=T_{\mu }^{\nu }
\end{equation}%
in which 
\begin{equation}
T_{\mu }^{\nu }=\frac{1}{2}\left( \mathcal{L}\delta _{\mu }^{\nu }-4\left(
F_{\mu \lambda }F^{\nu \lambda }\right) \frac{\partial \mathcal{L}}{\partial 
\mathcal{F}}\right) .
\end{equation}%
We note that $\mathcal{F}=2F_{tr}F^{tr}$ and therefore%
\begin{equation}
T_{t}^{t}=T_{r}^{r}=\frac{1}{2}\left( \mathcal{L}-2\mathcal{F}\frac{\partial 
\mathcal{L}}{\partial \mathcal{F}}\right) ,
\end{equation}%
while 
\begin{equation}
T_{\theta }^{\theta }=\frac{1}{2}\mathcal{L}.
\end{equation}%
We apply now the Legendre transformation \cite{3} $P_{\mu \nu }=\frac{%
\partial \mathcal{L}}{\partial \mathcal{F}}F_{\mu \nu }$ with $\mathcal{P}%
=P_{\mu \nu }P^{\mu \nu }=\left( \frac{\partial \mathcal{L}}{\partial 
\mathcal{F}}\right) ^{2}\mathcal{F}$ to introduce the Hamiltonian density as%
\begin{equation}
\mathcal{H}=2\mathcal{F}\frac{\partial \mathcal{L}}{\partial \mathcal{F}}-%
\mathcal{L}.
\end{equation}%
If one assumes that $\mathcal{H}=\mathcal{H}\left( \mathcal{P}\right) $ then
from the latter equation 
\begin{equation}
\frac{\partial \mathcal{H}}{\partial \mathcal{P}}d\mathcal{P}=\left( \frac{%
\partial \mathcal{L}}{\partial \mathcal{F}}+2\mathcal{F}\frac{\partial ^{2}%
\mathcal{L}}{\partial \mathcal{F}^{2}}\right) d\mathcal{F}
\end{equation}%
which implies%
\begin{equation}
\frac{\partial \mathcal{H}}{\partial \mathcal{P}}d\left( \left( \frac{%
\partial \mathcal{L}}{\partial \mathcal{F}}\right) ^{2}\mathcal{F}\right) =%
\frac{1}{\frac{\partial \mathcal{L}}{\partial \mathcal{F}}}\frac{\partial }{%
\partial \mathcal{F}}\left( \left( \frac{\partial \mathcal{L}}{\partial 
\mathcal{F}}\right) ^{2}\mathcal{F}\right) d\mathcal{F}
\end{equation}%
and consequently%
\begin{equation}
\frac{\partial \mathcal{H}}{\partial \mathcal{P}}=\frac{1}{\frac{\partial 
\mathcal{L}}{\partial \mathcal{F}}}.
\end{equation}%
Using the inverse transformation $F_{\mu \nu }=\frac{\partial \mathcal{H}}{%
\partial \mathcal{P}}P_{\mu \nu }$ one finds $\mathcal{F}=\left( \frac{%
\partial \mathcal{H}}{\partial \mathcal{P}}\right) ^{2}\mathcal{P}$ and
finally%
\begin{equation}
\mathcal{L}=2\mathcal{P}\frac{\partial \mathcal{H}}{\partial \mathcal{P}}-%
\mathcal{H}.
\end{equation}%
As a result of the Legendre transformation the Maxwell's equations become%
\begin{equation}
d\left( ^{\star }\mathbf{P}\right) =0
\end{equation}%
in which $\mathbf{P=}P_{\mu \nu }dx^{\mu }\wedge dx^{\nu }$ and 
\begin{equation}
T_{\mu }^{\nu }=\frac{1}{2}\left( \left( 2\mathcal{P}\frac{\partial \mathcal{%
H}}{\partial \mathcal{P}}-\mathcal{H}\right) \delta _{\mu }^{\nu }-4\left(
P_{\mu \lambda }P^{\nu \lambda }\right) \frac{\partial \mathcal{H}}{\partial 
\mathcal{P}}\right) .
\end{equation}%
From our field ansatz one easily finds that%
\begin{equation}
T_{t}^{t}=T_{r}^{r}=\frac{-\mathcal{H}}{2},
\end{equation}%
while 
\begin{equation}
T_{\theta }^{\theta }=\frac{1}{2}\left( 2\mathcal{P}\frac{\partial \mathcal{H%
}}{\partial \mathcal{P}}-\mathcal{H}\right) .
\end{equation}%
Let's choose now 
\begin{equation}
\mathcal{H}=\frac{2q^{2}\mathcal{P}}{s^{2}\mathcal{P}-2q^{2}}
\end{equation}%
in which $q$ and $s$ are two constants. Also from (3) we know that%
\begin{equation}
\mathbf{P=}D\left( r\right) dt\wedge dr
\end{equation}%
and therefore the Maxwell's equation (14) implies%
\begin{equation}
D\left( r\right) =\frac{Q}{r}.
\end{equation}%
Here $Q$ is an integration constant related to charge of the possible
solution. Having $D\left( r\right) $ available one finds $\mathcal{P}=-\frac{%
2Q^{2}}{r^{2}}$ and therefore 
\begin{equation}
\mathcal{H}=\frac{2Q^{2}}{s^{2}+r^{2}}
\end{equation}%
in which $Q=q$ is used. The $tt$ / $rr$ component of the Einstein's equation
with $G_{t}^{t}=G_{r}^{r}=\frac{A^{\prime }\left( r\right) }{2r}$ reads%
\begin{equation}
\frac{A^{\prime }\left( r\right) }{2r}+\Lambda =-\frac{Q^{2}}{\left(
s^{2}+r^{2}\right) }
\end{equation}%
for a prime denoting $\frac{d}{dr},$ which admits the following solution for
the metric function%
\begin{equation}
A\left( r\right) =C+\frac{r^{2}}{\ell ^{2}}-Q^{2}\ln \left(
r^{2}+s^{2}\right) ,
\end{equation}%
where $C$ is an integration constant and $\frac{1}{\ell ^{2}}=-\Lambda .$
The $\theta \theta $ component of energy momentum tensor is found to be%
\begin{equation}
T_{\theta }^{\theta }=\frac{Q^{2}\left( r^{2}-s^{2}\right) }{\left(
r^{2}+s^{2}\right) ^{2}}.
\end{equation}%
One can check that with $G_{\theta }^{\theta }=\frac{A^{\prime \prime
}\left( r\right) }{2}$ the $\theta \theta $ component of the Einstein
equations is also satisfied. Herein, without going through the detailed
calculations, we refer to the Brown and York formalism \cite{9} to show that 
$-C$ in (23) is the mass of the black hole i.e., $C=-M$. Such details in $%
2+1-$dimensions can also be found in Ref. \cite{10}. The asymptotic behavior
of the solution at large $r$ is the standard charged BTZ solution i.e., 
\begin{equation}
\lim_{r\rightarrow \infty }A\left( r\right) =-M+\frac{r^{2}}{\ell ^{2}}%
-2Q^{2}\ln r.
\end{equation}%
For small $r$ it behaves as 
\begin{equation}
\lim_{r\rightarrow 0}A\left( r\right) =-M-Q^{2}\ln s^{2}+\frac{r^{2}}{\ell
^{2}}
\end{equation}%
which makes the metric locally (anti-)de Sitter. Furthermore, one observes
that all invariants are finite at any $r\geq 0$ \cite{1}. Next, explicit
form of the Lagrangian density with respect to $\mathcal{P}$ is given by%
\begin{equation}
\mathcal{L}=\frac{-2Q^{2}\mathcal{P}\left( 2Q^{2}+s^{2}\mathcal{P}\right) }{%
\left( 2Q^{2}-s^{2}\mathcal{P}\right) ^{2}}
\end{equation}%
and the closed form of the electric field i.e., $E\left( r\right) =\frac{%
\partial \mathcal{H}}{\partial \mathcal{P}}D\left( r\right) $ becomes%
\begin{equation}
E\left( r\right) =-\frac{Qr^{3}}{\left( s^{2}+r^{2}\right) ^{2}}.
\end{equation}%
We comment that $E\left( r\right) $ is also regular everywhere and at large $%
r$ it behaves similar to the standard linear Maxwell's field theory. In all
our results, setting $s$ to zero takes our solution to the standard charged
BTZ solution. We must add that the metric function (23) provides a regular
solution. Depending on the parameters $M,$ (or $C$), $Q$ and $s$ it may give
a black hole with single / double horizon, or no horizon at all (See Fig.s 1
and 2). 

\subsection{Energy Conditions and Thermodynamics in brief}

In this part we would like to check the energy conditions such as the weak
(WECs) and the strong energy conditions (SECs). As we have found, the energy
density is given by%
\begin{equation}
\rho =-T_{t}^{t}=\frac{Q^{2}}{\left( s^{2}+r^{2}\right) }
\end{equation}%
while the radial and tangential pressures are given respectively by%
\begin{equation}
p_{r}=T_{r}^{r}=-\frac{Q^{2}}{\left( s^{2}+r^{2}\right) }
\end{equation}%
and%
\begin{equation}
p_{\theta }=T_{\theta }^{\theta }=\frac{Q^{2}\left( r^{2}-s^{2}\right) }{%
\left( r^{2}+s^{2}\right) ^{2}}.
\end{equation}%
WECs imply i) $\rho \geq 0$ ii), $\rho +p_{r}\geq 0$ and iii) $\rho
+p_{\theta }\geq 0.$ All of these conditions are trivially satisfied. The
SECs state that in addition to WECs we must also have $\rho +p_{r}+p_{\theta
}\geq 0$ leading to the condition%
\begin{equation}
\frac{Q^{2}\left( r^{2}-s^{2}\right) }{\left( r^{2}+s^{2}\right) ^{2}}\geq 0
\end{equation}%
which is satisfied for $r\geq s$. In conclusion WECs are satisfied
everywhere while SECs are satisfied only for $r\geq s$.

To complete our solution we look at the thermodynamics of the solution. (A
comprehensive study on thermodynamics of Einstein-Born-Infeld black holes in
three dimensions can be found in Ref. \cite{11}). If we consider $r_{h}$ to
be the radius of the event horizon then%
\begin{equation}
A\left( r_{h}\right) =0
\end{equation}%
which implies the mass given by%
\begin{equation}
M=\frac{r_{h}^{2}}{\ell ^{2}}-Q^{2}\ln \left( r_{h}^{2}+s^{2}\right) .
\end{equation}%
From the first law of thermodynamics $dM=TdS+\Phi dQ$ in which $S=2\pi r_{h}$
is the entropy and $\Phi $ is the electric potential all measured at the
horizon, one can write%
\begin{equation}
T=\left( \frac{\partial M}{\partial S}\right) _{Q}=\frac{r_{h}\left(
r_{h}^{2}+s^{2}-Q^{2}\ell ^{2}\right) }{2\pi \ell ^{2}\left(
r_{h}^{2}+s^{2}\right) }.
\end{equation}%
Finally we write the heat capacity as $C_{Q}=T\left( \frac{\partial T}{%
\partial S}\right) $ which is given by%
\begin{equation}
C_{Q}=\frac{r_{h}\left( r_{h}^{2}+s^{2}-Q^{2}\ell ^{2}\right) \left(
r_{h}^{4}+r_{h}^{2}\left( 2s^{2}+Q^{2}\ell ^{2}\right) +s^{2}\left(
s^{2}-\ell ^{2}Q^{2}\right) \right) }{16\pi ^{3}\ell ^{4}\left(
r_{h}^{2}+s^{2}\right) ^{3}}.
\end{equation}%
One observes that $\lim_{s\rightarrow 0}$ $T=\frac{r_{h}^{2}-Q^{2}\ell ^{2}}{%
2\pi \ell ^{2}r_{h}}$ and $\lim_{s\rightarrow 0}$ $C_{Q}=\frac{%
r_{h}^{4}-Q^{4}\ell ^{4}}{16\pi ^{3}\ell ^{4}r_{h}^{3}}$ which are the
thermodynamic quantities of charged BTZ.

In brief, we rederived the $2+1-$dimensional version of the regular Bardeen
black hole. Our source is NED with an electric field $F_{tr}\neq 0,$ in $%
2+1- $dimensions. Our Maxwell invariant $\mathcal{F}=F_{\mu \nu }F^{\mu \nu
} $ is regular everywhere. For $r\rightarrow \infty $ our solution goes to
the charged BTZ solution. For $r\rightarrow 0$ the solution is locally
(anti)-de Sitter which globally can be interpreted as a topological defect.

\section{Charge Screening by a Thin Stable Shell}

In this section we shall use the formalism introduced by Eiroa and Simeone 
\cite{7} to construct a thin-shell (not bubble) which may shield the charge
of the regular Bardeen black hole given above. (There are some other related
works in $2+1-$dimensions which are given in Ref. \cite{12}) Therefore we
employ the Bardeen black hole solution in $2+1-$dimensions for $r<a$ (region
1 with $f_{1}\left( r\right) =A\left( r\right) $ in (2)) and the de Sitter
BTZ black hole solution for $r>a$ (region 2 with $f_{2}\left( r\right)
=A\left( r\right) $ in (2)) in which $a$ is the radius of the thin-shell
under construction. The extrinsic line element on the shell (or ring) is
written as 
\begin{equation}
ds_{12}^{2}=-d\tau ^{2}+a^{2}d\theta ^{2}.
\end{equation}%
where $\tau $ is the proper time on our timelike shell. One must note that
our shell is not dynamic in general but in order to investigate the
stability of the thin shell against a linear perturbation, we let the radius 
$a$ to be a function of the proper time $\tau $ which is measured by an
observer on the shell. This indeed does not mean that the bulk metric is
time dependent. This method has been introduced by Israel \cite{5} and being
used widely to study the stability of thin-shell and thin-shell wormholes
ever since \cite{13}. The Einstein equations on the shell become Lanczos
equations \cite{5,7} which are given by%
\begin{equation}
-[K_{i}^{j}]+\left[ K\right] \delta _{i}^{j}=8\pi S_{i}^{j}
\end{equation}%
in which one finds \cite{7} the energy density on the shell 
\begin{equation}
\sigma =-S_{\tau }^{\tau }=\frac{1}{8\pi a}\left( \sqrt{f_{1}\left( a\right)
+\dot{a}^{2}}-\sqrt{f_{2}\left( a\right) +\dot{a}^{2}}\right) 
\end{equation}%
and the pressure 
\begin{equation}
p=S_{\theta }^{\theta }=\frac{2\ddot{a}+f_{2}^{\prime }\left( a\right) }{%
16\pi \sqrt{f_{2}\left( a\right) +\dot{a}^{2}}}-\frac{2\ddot{a}%
+f_{1}^{\prime }\left( a\right) }{16\pi \sqrt{f_{1}\left( a\right) +\dot{a}%
^{2}}}.
\end{equation}%
Note that a 'prime' is derivative wrt $a$ while a 'dot' is wrt proper time.
Having energy conserved implies that 
\begin{equation}
\frac{d}{d\tau }\left( a\sigma \right) +p\frac{da}{d\tau }=0
\end{equation}%
for any dynamic shell (ring) as $a$ is a function of proper time $\tau .$ If
one considers the equilibrium radius to be at $a=a_{0}$ the energy density
and pressure at equilibrium are given by%
\begin{equation}
\sigma _{0}=\frac{1}{8\pi a_{0}}\left( \sqrt{f_{1}\left( a_{0}\right) }-%
\sqrt{f_{2}\left( a_{0}\right) }\right) 
\end{equation}%
and%
\begin{equation}
p_{0}=\frac{f_{2}^{\prime }\left( a_{0}\right) }{16\pi \sqrt{f_{2}\left(
a_{0}\right) }}-\frac{f_{1}^{\prime }\left( a_{0}\right) }{16\pi \sqrt{%
f_{1}\left( a_{0}\right) }}.
\end{equation}%
Furthermore a linear perturbation causes the pressure and energy density to
vary as $p\simeq p_{0}+\beta \sigma $ in which $\beta $ is a parameter
equivalent to the speed of sound on the shell. Next, one can, in principle,
solve the conservation of energy equation to find 
\begin{figure}[tbp]
\includegraphics[width=130mm,scale=1]{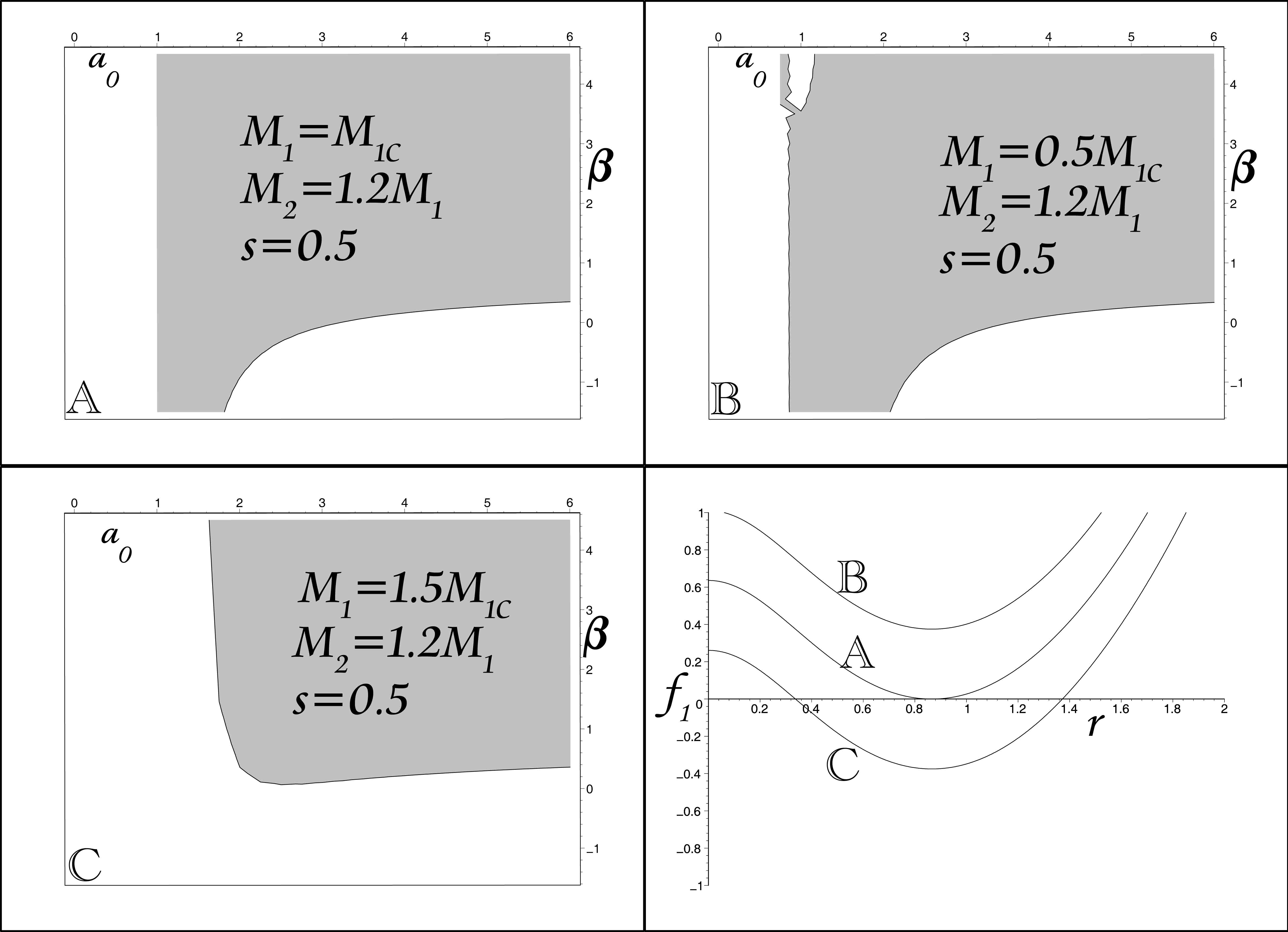}
\caption{A plot of $V^{\prime \prime }\left( a_{0}\right) $ versus $\protect%
\beta $ and $a_{0}$ with $Q=1,\ell ^{2}=1$ and $M_{1}=M_{1c}$ in (A) $%
M_{1}=0.5M_{1c}$ in (B) and $M_{1}=1.5M_{1c}$ in (C). For all three plots $%
M_{2}=1.2M_{1}$ and $s=0.5.$ In the right bottom we also plot the metric
function for $r<a_{0}$ to show that the possible horizon remains inside the
thin shell. Figures (A) and (B) manifest stability for the thin shell
against a linear perturbation. }
\end{figure}
\begin{figure}[tbp]
\includegraphics[width=130mm,scale=1]{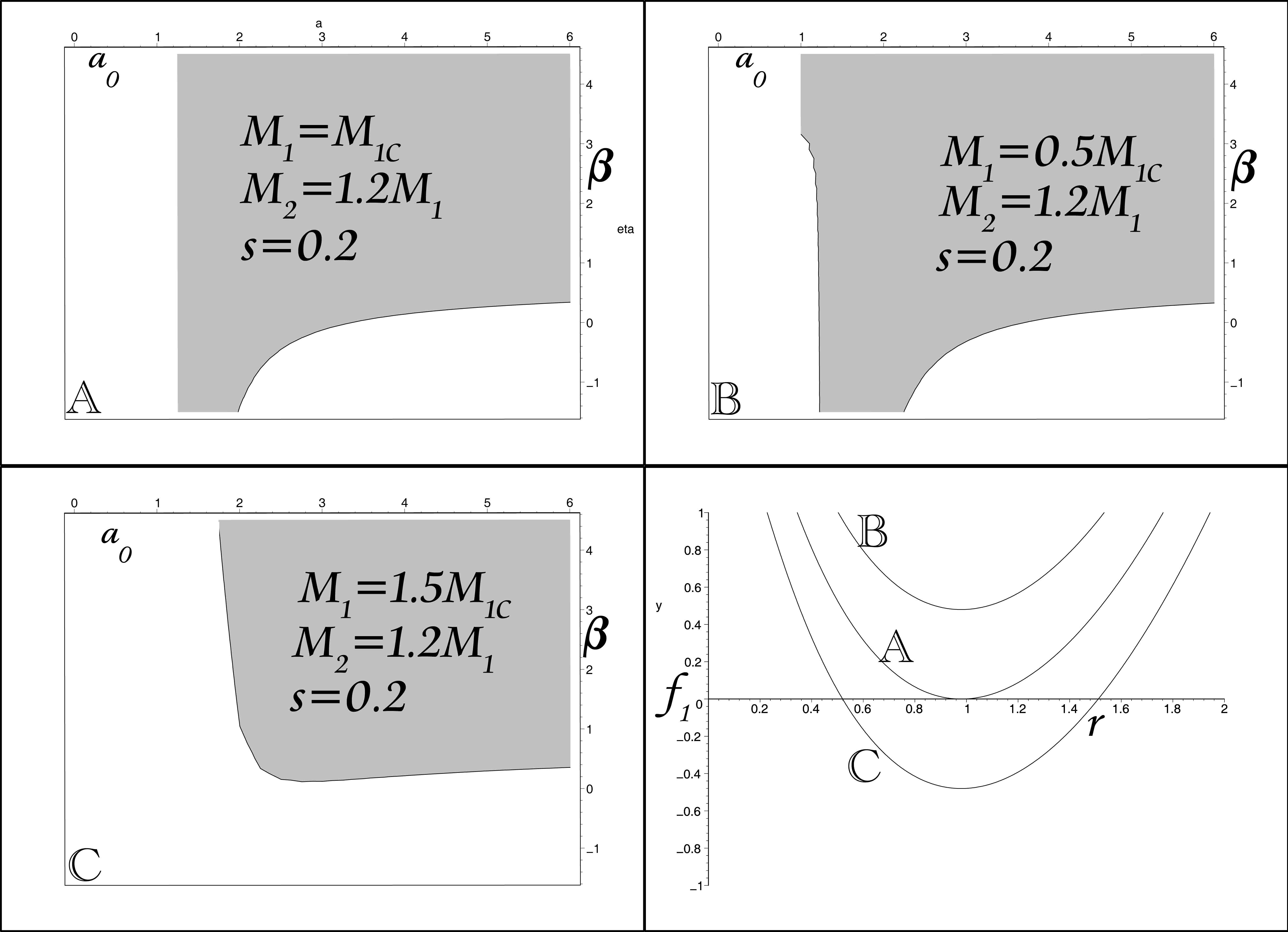}
\caption{A plot of $V^{\prime \prime }\left( a_{0}\right) $ versus $\protect%
\beta $ and $a_{0}$ with $Q=1,\ell ^{2}=1$ and $M_{1}=M_{1c}$ in (A) $%
M_{1}=0.5M_{1c}$ in (B) and $M_{1}=1.5M_{1c}$ in (C). For all three plots $%
M_{2}=1.2M_{1}$ and $s=0.2.$ In the right bottom we also plot the metric
function for $r<a_{0}$ to show that the possible horizon remains inside the
thin shell. Figures (A) and (B) manifest stability for the thin shell
against a linear perturbation.}
\end{figure}
\begin{equation}
\sigma =\left( \frac{a}{a_{0}}\right) ^{1+\beta }\left( \sigma _{0}+\frac{%
p_{0}}{1+\beta }\right) -\frac{p_{0}}{1+\beta }.
\end{equation}%
The dynamic of the energy density also is given by (39) which together imply
a one dimensional equation of motion for the shell is given by%
\begin{equation}
\dot{a}^{2}+V\left( a\right) =0
\end{equation}%
with 
\begin{equation}
V\left( a\right) =\frac{f_{1}\left( a\right) +f_{2}\left( a\right) }{2}%
-\left( \frac{f_{1}\left( a\right) -f_{2}\left( a\right) }{16\pi a\sigma }%
\right) ^{2}-\left( 4\pi a\sigma \right) ^{2}.
\end{equation}%
At the equilibrium $V\left( a_{0}\right) =V^{\prime }\left( a_{0}\right) =0$
and the first nonzero term is the second derivative of the potential at $%
a=a_{0}$ which must be positive to have an oscillatory motion for the shell
upon linear perturbation. This in turn means that the shell will be stable.
In Fig.s 1 and 2 we plot the region in which $V^{\prime \prime }\left(
a_{0}\right) \geq 0$ or the stable regions with 
\begin{equation}
f_{1}\left( a\right) =-M_{1}+\frac{a^{2}}{\ell ^{2}}-Q_{1}^{2}\ln \left(
a^{2}+s^{2}\right) 
\end{equation}%
and 
\begin{equation}
f_{2}\left( a\right) =-M_{2}+\frac{a^{2}}{\ell ^{2}}.
\end{equation}%
To do so we used a critical mass 
\begin{equation}
M_{1c}=Q^{2}\left[ 1-\ln \left( Q^{2}\ell ^{2}\right) \right] -\frac{s^{2}}{%
\ell ^{2}}
\end{equation}%
at which for $M_{1}>M_{1c}$ a black hole with two horizons forms inside the
thin-shell and for $M_{1}<M_{1c}$ the solution for inside thin-shell is
non-black hole while $M_{1}=M_{1c}$ represents the extremal black hole
inside the thin-shell. We note that a distant observer does not detect any
electric charge of the black hole. Therefore the black hole structure of the
spacetime inside the shell may not be seen even though $a$ was supposed to
be larger than the horizon. In this case the thin-shell carries a charge $%
Q_{2}=-Q_{1}$ which completely shields the black hole nature of the
spacetime. In fact Fig.s 1 and 2 show that the thin-shell is stable for all
values of $\beta $ irrespective of whether we have an extremal black hole or
no black hole at all. However in the case of non-black hole solution which
are shown in Fig.s 1B and 2B, if the initial radius of the ring $a_{0}$
(which is also the equilibrium radius) is set less than $r_{\min }$ in which 
$f_{1}^{\prime }\left( r_{\min }\right) =0$, such perturbation may make the
ring to collapse to a point. In such case still there is no singularity and
the remained spacetime is BTZ solution. Furthermore, since the internal
black hole is regular, the thin-shell behaves like an ordinary object with
no singularity inside. 

\section{Conclusion}

No doubt, Einstein / Einstein-Maxwell theory has limited scopes in $2+1-$%
dimensional spacetimes which has been studied extensively during the recent
decade \cite{8}. With non-linear electrodynamics (NED) fresh ideas has been
pumped into the spacetime and served well as far as removal of singularities
is concerned \cite{10,14}. Most of the black hole properties in $3+1-$%
dimensional spacetime has counterparts in $2+1-$dimensions with yet some
differences. One common property is the existence of regular Bardeen black
hole which is sourced by a radial electric field in $2+1-$dimensions whereas
the source in $3+1-$dimensions happened to be magnetic. By encircling the
central regular Bardeen black hole by a charged thin-shell and matching
inside to outside in accordance with the Lanczos' conditions we erase the
entire effect of charge to the outside world. Such a thin-shell (or ring)
doesn't seem a mere illusion, but is a reality since it turns out to be
stable against linear perturbations. The idea works in the case of a regular
interior black hole well but remains to be proved whether it is applicable
for a singular black hole. From astrophysical point of view is it possible
that a natural, concentric thin-shell with equal (but opposite) charge to
that of a central black hole forms to cancel external effect of charge at
all? Admittedly our analysis here relates only to the $2+1-$dimensional case
but it is natural to expect a similar charge screening effect in higher
dimensions as well.

\end{document}